\documentstyle[preprint,revtex,eqsecnum]{aps}

\begin{document}
\draft
\begin{title}
\begin{center}
Absence of Dipole Transitions in Vortices of
Type II Superconductors
\end{center}
\end{title}
\author{Theodore C. Hsu\cite{byline}}
\begin{instit}
Department of Physics, University of British Columbia\\
Vancouver, B.C. Canada, V6T 1Z1
\end{instit}
\receipt{}
\begin{abstract}
The response of a single vortex to a time
dependent field is examined microscopically and
an equation of motion for vortex motion at
non-zero frequencies is derived.
Of interest are frequencies
near $\Delta^{2}/E_{F}$, where $\Delta$ is the
bulk energy gap and $E_{F}$ is the fermi energy.
The low temperature, clean,
extreme type II limit and maintaining of equilibrium with the lattice
are assumed. A simplification occurs
for large planar mass anisotropy. Thus the results may be
pertinent to materials such as $NbSe_2$ and
high temperature superconductors.
The expected dipole transition between
core states is hidden
because of the self consistent
nature of the vortex potential. Instead the vortex
itself moves and has a resonance at the frequency of the
transition.
\end{abstract}
\narrowtext

The description of quasiparticle levels
in vortex cores
of superconductors has been known
for some time. Caroli {\it et al.}\cite{CAROLIA}
and Bardeen {\it et al.}\cite{BARDEEN}
calculated the energies and
wavefunctions of these discrete levels
using the Bogoliubov-deGennes (BdG) equation.
This was a basis
for theories of dissipative vortex motion
based on the idea of a `normal' core such as
those of Bardeen and Stephen\cite{BARDEEN_STEPHEN},
Nozieres and Vinen (NV)\cite{NOZIERES} and others\cite{OTHERS}.
Kramer and Pesch\cite{KRAMER}
used the
Eilenberger equation
to calculate
the density of states in the core.
Their approach has proven useful in
qualitatively explaining\cite{THEORY}
scanning tunneling microscope experiments on
$NbSe_{2}$ by Hess {\it et al.}\cite{HESS}.
These experiments, however, probe the density of states at
a scale $ \sim 0.1 meV$ whereas the separation
of core levels is $\sim 10 mK$.
Caroli and Matricon\cite{CAROLIB} discussed the implications of
discrete levels for ultrasound attenuation
and nuclear magnetic relaxation. Transitions between
these levels may have been observed recently in high temperature
superconductors \cite{Karrai}. The electromagnetic
response is interesting from a practical point of view.
Herein we focus on the low temperature and clean limit, considering
eigenfunctions and matrix elements of
quasiparticle states, and assuming
the BdG
equations and a local gap equation,
$\Delta \left({\vec r}\right) =
V\langle c_{\uparrow}\left({\vec r}\right)
c_{\downarrow}\left({\vec r}\right)\rangle$,
($c_{\sigma}$ is a spin $\sigma$ electron operator)
to be valid. The vortex response to an electromagnetic
field will be considered from a purely microscopic
point of view. Real materials have vortex pinning
but at frequencies of order of magnitude comparable
to the core level separation undergo a cross-over
to unpinned behaviour\cite{GITTLEMAN} so a first step
should be a study of
unpinned vortices at those frequencies.
Real superconductors are
non-local and the interaction is retarded
but the hope is that, since we deal with rigid
motions and relatively low frequencies,
some relevance to real materials remains.

In terms of eigenfunctions,
$\psi_{\mu}({\vec r})^{T} = \left(
u_{\mu}({\vec r})\quad
v_{\mu}({\vec r})
\right)$,
the quasiparticle operators are,
\begin{equation}
\left(
\begin{array}{c}
\gamma_{\mu\uparrow}^{\dagger}\\
\gamma_{\mu\downarrow}^{\dagger}
\end{array}
\right)
=\int d^{3}{\vec r}
\left(
\begin{array}{cc}
c_{\uparrow}^{\dagger}({\vec r})
& c_{\downarrow}({\vec r})\\
c_{\downarrow}^{\dagger}({\vec r})
&-c_{\uparrow}({\vec r})
\end{array}
\right)\psi_{\mu}({\vec r}).
\label{QUASIPARTICLE}
\end{equation}
The Schr\"odinger equation for $\psi$ is the BdG equation,
\begin{equation}
\epsilon\psi ({\vec r}) = \sigma^{z}
\left[{1\over 2m}
({\vec p} - \sigma^{z}{e\over c}{\vec A})^{2}
- E_{F}\right]\psi ({\vec r}) +
\left(
\begin{array}{cc}
0&\Delta ({\vec r})\\
\Delta^{*}({\vec r}) &0
\end{array}
\right)\psi ({\vec r})
\label{BGD}
\end{equation}
where $\Delta = |\Delta({\vec r}-{\vec r_{0}})|
\exp{(-i\theta({\vec r}-{\vec r_{0}}))}$.
$\theta({\vec r}-{\vec r_{0}})$ is the angle about the
center of the vortex ${\vec r_{0}}$
measured from the ${\hat x}$ axis. We consider a
vortex parallel to ${\hat z}$.

In the extreme type II limit with $H << H_{c2}$
the magnetic field
creating vortices may be ignored.
Its importance compared to the phase of $\Delta$
is reduced by $\xi^{2}/\lambda^{2}$ where
$\xi$ is the coherence length and $\lambda$ is the
penetration depth.

The eigenfunctions for fixed $k_{z}$, $\mu << k_{F\perp}\xi$,
and the radial coordinate $r << \xi$ are
\begin{equation}
\psi_{\mu}({\vec r}) = \left({k_{F}\over{2\pi\xi L_{z}}}\right)^{2}
e^{ik_{z}z}
\left(
\begin{array}{c}
e^{i(\mu - {1\over 2})\phi}J_{\mu - {1\over 2}}(k_{F\perp}r)\\
e^{i(\mu + {1\over 2})\phi}J_{\mu + {1\over 2}}(k_{F\perp}r)
\end{array}
\right)
\label{EIGENFUNCTIONS}
\end{equation}
where $\mu = \pm{1\over 2}, \pm{3\over 2}, ...$
and $\perp$ refers to the ${\hat x},{\hat y}$ directions
(we assume at least cylindrical symmetry). They fall off
exponentially for $r > \xi$.
The energies as calculated by Kramer and Pesch\cite{KRAMER},
who accounted for some self-consistency effects
due to the gap equation, are
\begin{equation}
\epsilon_{\mu} =
{
{2\mu \Delta_{0}^{2}}
\over
{k_{F}v_{F}cos^{2}\Theta}
}
ln({\pi\over 2}\xi_{0}cos\Theta/\xi_{1}),\quad cos\Theta\equiv
k_{F\perp}/k_{F}.
\label{EIGENVALUE}
\end{equation}
The logarithmic factor is
not important here and ignored hereafter but see reference \cite{LARKIN}.

Consider a
long wavelength electromagnetic
wave, ${\vec A}^{\prime} \perp {\hat z}$,
with polarization at angle $\theta_{0}$
to ${\hat x}$. We treat the perturbation
$-{e\over{mc}}{\vec A}^{\prime}\cdot{\vec p}$
to first order in $A^{\prime}$.
If the matrix elements with respect to low energy states
are not very sensitive to
the exact behaviour of the wavefunctions at the core
boundary
then they
may be estimated,
using standard Bessel function identities, to be
$\int \psi_{\mu\pm 1}^{\dagger}
(-{e\over{mc}}{\vec A}^{\prime}\cdot{\vec p})
\psi_{\mu}
= (
{e\hbar k_{F\perp}A^{\prime}}
/
{2mc}
)
\exp{\mp i(\theta_{0} + {\pi\over 2})}
$.

Consider a
vortex with velocity ${\vec v}_{L} \perp {\hat z}$
a background superfluid velocity
${\vec v}_{S}$.
These velocities will be assumed uniform along the length of the vortex.
This is valid if the distance the electromagnetic
wave penetrates the superconductor (the shorter of the London penetration
depth and skin depth) is long
compared to the coherence length.
We shall consider a background supercurrent or
gauge field of the form,
\begin{equation}
{\vec A}^{\prime}(t)
\equiv
-(mc/e){\vec v}_{S}(t)
=
{\vec E}ct
+ {\vec A}^{\prime}_{0}.
\label{AFIELD}
\end{equation}
${\vec E}$ is the
applied electric field.

Let us now outline the ensuing calculation.
The time dependence of the
quasiparticle states under
the applied field and the moving vortex
will be calculated
and written in terms of a density matrix.
The motion of the vortex will be inferred, using
the gap equation, by
identifying changes in the off-diagonal
components of the density matrix
with that due to a displaced vortex.
Because this calculation involves
self-consistency in the vortex velocity
there is a slight subtlety.
Given inital (t=0)
values of
${\vec v}_{L}$ and
${\vec v}_{S}$, as considered below,
in general they will not be consistent.
In general,
${\vec v}_{L}$ is
obtained by integrating the equation of motion
for a given
${\vec v}_{S}(t)$.
We shall first derive the acceleration of the
vortex at $t=0$ and then find the equation of motion for
all time by
beginning in a well defined equilibrium state (
${\vec v}_{L} = 0$,
${\vec v}_{S} = 0$,
${\vec r}_{0} = 0$ at $t=0$)
and, given a time dependence
${\vec v}_{S}(t)$,
calculating the resulting motion to all orders in t.

Suppose the field in equation \ref{AFIELD}
is turned on at $t=0$.
Let us calculate the quasiparticle
density matrix to order $t^{2}$.
We use $\langle\rangle_{0}$ to denote the
(diagonal) values at $t=0$.
Using matrix elements of ${\vec A}^{\prime}$
the off-diagonal density matrix elements are,
\begin{equation}
\langle \gamma_{\mu}^{\dagger}\gamma_{\mu - 1}\rangle(t)
=
[\langle \gamma_{\mu}^{\dagger}\gamma_{\mu}\rangle_{0}
-
\langle \gamma_{\mu - 1}^{\dagger}\gamma_{\mu - 1}\rangle_{0}
]
\times
\left\{-i{W\over\hbar}t - {t^{2}\over{2\hbar^{2}}}
\left[
i\hbar W^{\prime} + W(\epsilon_{\mu - 1} - \epsilon_{\mu})
\right]\right\}
\label{OFFDIAGONAL}
\end{equation}
where $W = (e\hbar k_{F\perp}A^{\prime}_{0}/2mc)
\exp{i(\theta_{0} + {\pi\over 2})}$
and $W^{\prime}$ is the same as W except with $A^{\prime}_{0}$
replaced by $cE$. Terms of order
$W^{2}$ (such as the change in the diagonal density matrix element)
are ignored by taking the amplitude of the perturbation to be sufficiently
small.

The significance of these density matrix elements
is clarified
by considering a displaced vortex
in terms of
the underlying quasiparticles.
The inverse of equation
\ref{QUASIPARTICLE} substituted into
the gap equation is
\begin{equation}
\Delta({\vec r}) = V\sum_{\mu ,\nu}
[(\delta_{\mu\nu} - \sum_{\sigma}
\langle \gamma_{\nu\sigma}^{\dagger}\gamma_{\mu\sigma}\rangle)
v_{\nu}^{*}({\vec r}) u_{\mu}({\vec r})
+ {\rm other\ terms}
].
\end{equation}
Suppose the quasiparticles are displaced by
$\delta {\vec r}_{0}$ at
angle $\phi_{0}$ to ${\hat x}$. The occupation
is taken to be diagonal before displacement,
$\langle \gamma_{\nu\sigma}^{\dagger}\gamma_{\mu\sigma}\rangle
=
\delta_{\mu\nu}f(\epsilon_{\nu})$, where $f(\epsilon)$ is
the fermi function. Then, in the undisplaced
eigen-basis, to linear order in $\delta r_{0}$,
\begin{eqnarray}
\delta\Delta ({\vec r}) =
-V\delta r_{0} k_{F\perp}
\sum_{\nu}
&[e^{i\phi_{0}}(f(\epsilon_{\nu}) - f(\epsilon_{\nu + 1}))
v_{\nu + 1}^{*}({\vec r}) u_{\nu}({\vec r})\\
&+
e^{-i\phi_{0}}(f(\epsilon_{\nu}) - f(\epsilon_{\nu + 1}))
v_{\nu}^{*}({\vec r}) u_{\nu + 1}({\vec r})
].
\label{DELTA}
\end{eqnarray}

Comparing with equation \ref{OFFDIAGONAL}
it is clear that the term of order t may be produced
by a rigid
translation of the vortex core at velocity
$\delta r_{0}/t = -(e/mc)A^{\prime}_{0}$
in direction $\phi_{0} = \theta_{0}$. It is
simply the velocity by which the gauge field boosts the group velocity
of all waves.

Returning to equation \ref{OFFDIAGONAL} the first
term of order $t^{2}$
corresponds to an acceleration
proportional to the electric field and equation \ref{DELTA}
shows that it is
$\dot{{\vec v}}_{S}$.
The second piece is trickier.
Because
$\epsilon_{\mu - 1} - \epsilon_{\mu}$ can depend on $k_{z}$
and (for large $\mu$) $\mu$,
the $t^{2}$ term does not correspond to a rigid acceleration of
the vortex core except in the limit of low temperature
and strong mass anisotropy,
$m_{z} >> m_{\perp}$.
There
$\epsilon_{\mu - 1} - \epsilon_{\mu} = 2\Delta_{0}^{2}/k_{F}v_{F}$
(independent of $\mu$) and making the identification
with a rigid displacement leads to an
acceleration
\begin{equation}
{{2\delta r_{0}}\over{t^{2}}}
=
-{
{2\Delta_{0}^{2}eA^{\prime}_{0}}\over{\hbar k_{F}v_{F}mc}
}
\label{LORENTZ}
\end{equation}
in a direction $\phi_{0} = \theta_{0} + {\pi\over 2}$
perpendicular to ${\vec v}_{S}$. This corresponds to
the Lorentz force.

At high temperatures higher energy levels become important.
The level spacing decreases with increasing energy
so higher energy quasiparticles
have a smaller acceleration
and lag behind the core. The term
$\epsilon_{\mu - 1} - \epsilon_{\mu}$
contains a factor $(1/cos^{2}\Theta)$
which is the $k_{z}$ dependence.
This factor is not actually divergent.
The expression \ref{EIGENVALUE} for the energies
is valid only for small energies.
The $\epsilon_{\mu}$ are bounded by $\Delta_{0}$ and
so $\epsilon_{\mu - 1} - \epsilon_{\mu}$
must go to zero as $\Theta \rightarrow {\pi /2}$.
In real materials such as $NbSe_{2}$ the fermi surface is open
which restricts $cos\Theta$.
Below, for simplicity, we assume that the
system has an anisotropic mass and that the lack of rigid
acceleration is not important.

Assuming rigid motion we see that
the applied field, instead of causing
dipole transitions, causes the density matrix
to evolve off-diagonal elements
corresponding to vortex motion
(after applying the gap equation).
In the new displaced set of basis functions the
density matrix is again diagonal. The vortex
does not stand still and allow a dipole transition
to take place, as the core of an atom.
{\it The vortex
is a self consistent potential}.

The moving vortex itself affects the
density matrix.
Suppose the vortex has velocity ${\vec v}_{L}$
at an angle $\phi_{0}$ to ${\hat x}$.
The matrix element
\begin{equation}
W_{\mu\nu} = \int d^{3}{\vec r} ~\psi_{\mu}^{\dagger}({\vec r})
\left(
\begin{array}{cc}
0 &\delta\Delta (t)\\
\delta\Delta^{*}(t) &0
\end{array}
\right)
\psi_{\nu}({\vec r})
\end{equation}
may be re-written using Galilean invariance.
Let $\delta\psi_{\mu}({\vec r})$ be the change in
$\psi_{\mu}({\vec r})$ upon displacement of the vortex
by $\delta{\vec r}_{0}$.
Then substituting $\delta\psi$ and $\delta\Delta$ into
the BdG equation, subtracting the undisplaced
piece, keeping terms to first order in $\delta r_{0}$
and integrating by parts results in
$W_{\mu\nu} = (\epsilon_{\nu} - \epsilon_{\mu})
\int \psi_{\mu}^{\dagger}({\vec r})
\delta\psi_{\nu}({\vec r})$.
Writing $\delta\psi ({\vec r}) =
t{\vec v}_{L}\cdot {\vec\nabla}\psi ({\vec r})$,
and using standard identities,
\begin{equation}
W_{\mu\nu} = -{
{{\rm v_{L}}tk_{F\perp}}
\over
{2i}
}
\delta_{\mu ,\nu\mp 1}
(\epsilon_{\nu} - \epsilon_{\mu})
e^{\pm i(\phi_{0} + {\pi\over 2})}.
\end{equation}
Treating this to linear order and
integrating from 0 to t gives an acceleration
$({2\delta r_{0}}/{t^{2}})
=
({\rm v_{L}}/\hbar)
(\epsilon_{\nu} - \epsilon_{\nu + 1})
=
-({\Delta_{0}^{2}}/{E_{F}\hbar}){\rm v_{L}}$
in a direction at an angle $+\pi/2$ to ${\vec v}_{L}$.

This together with equation \ref{LORENTZ}
gives an acceleration
$(\Delta_{0}^{2}/\hbar E_{F}) ({\vec v}_{L} - {\vec v}_{S}) \times {\hat z}$
corresponding to the Magnus force given in reference \cite{NOZIERES}
as,
$ (hn/2)({\vec v}_{S}-{\vec v}_{L})\times{\hat z}$,
where $n$ is the (superfluid) electron density.
Taking
$n=k_{F\perp}^{2}k_{Fz}/\pi^{3}$ and
the in-plane coherence length
$\xi_{\perp} = \hbar v_{F\perp}/\pi\Delta$,
one may extract a `mass' of the vortex
$M \sim m_{\perp}(k_{F\perp}\xi_{\perp})^{2}$
per unit length $k_{Fz}^{-1}$.
This expression is perhaps a microscopic justification
for a `normal core' of size
$\xi$, even at low temperatures when there
is a gap in the single particle density of states.
This mass should be contrasted with the very different
definition of mass discussed recently by Duan and Leggett \cite{Duan}.
Here the mass corresponds to the inertia of the electrons in the core
of the vortex.

To treat dissipation
we assume that core states
maintain equilibrium with the lattice
and that the linear in t change in the single particle states
(in the lattice frame of reference) also decays as an exponential
$\exp{(-t/\tau )}$.
$\tau$ is
related to the transport lifetime although there are differences
in matrix element and phase space which can be
elucidated in a microscopic theory.
Given the velocity ${\vec v}_{L}$, then to linear order
in t the off-diagonal component of the density matrix is,
in a basis fixed to the lattice at $t=0$,
\begin{eqnarray}
\langle \gamma_{\mu}^{\dagger}\gamma_{\mu -1}\rangle
&=
\langle \gamma_{\mu}^{\dagger}\gamma_{\mu}\rangle
(1-
\langle \gamma_{\mu -1}^{\dagger}\gamma_{\mu -1}\rangle)
[-({k_{F\perp}v_{L}}/{2})e^{i\phi_{0}}]t\\
&+
\langle \gamma_{\mu -1}^{\dagger}\gamma_{\mu -1}\rangle
(1-
\langle \gamma_{\mu}^{\dagger}\gamma_{\mu}\rangle)
[({k_{F\perp}v_{L}}/{2})e^{i\phi_{0}}]t
\end{eqnarray}
The result is an additional
contribution to the second derivative of the off-diagonal element,
\begin{equation}
{{d^{2}}\over{dt^{2}}}
\langle \gamma_{\mu}^{\dagger}\gamma_{\mu -1}\rangle
=
-{1\over\tau}{d\over{dt}}
\langle \gamma_{\mu}^{\dagger}\gamma_{\mu -1}\rangle
+ ({\rm previous\ terms}).
\end{equation}
This produces a contribution $-(1/\tau ){\vec v}_{L}$ to the
vortex acceleration.
Collecting terms the equation of motion is
\begin{equation}
\dot{{\vec v}}_{L}
=
\dot{{\vec v}}_{S}
+
{{\Delta_{0}^{2}}\over
{\hbar E_{F}}}
({\vec v}_{L}- {\vec v}_{S})\times{\hat z}
-
{1\over\tau}
{\vec v}_{L}.
\label{MOTION}
\end{equation}

In the above derivation
we applied ${\vec v}_{S}$ instantaneously
to a vortex stationary for $t < 0$. That produced
${\vec v}_{L} = {\vec v}_{S}$ at $t=0$ which
is an insufficiently general initial condition.
We now show that equation \ref{MOTION}
is valid at all times. We begin at $t=0$
in a well defined state,
${\vec v}_{L} = {\vec v}_{S} = 0$, and displacement ${\vec r}_{0} = 0$.
Given the Taylor expansion of ${\vec v}_{S}$
we may calculate, using the above technique,
the displacement to any order $t^{n}$
because we need know only
${\vec v}_{S}$ to order $n-1$ and ${\vec v}_{L}$ to order
$n-2$. The gap equation applied to the order $n$
displacement of the quasiparticles
gives ${\vec v}_{L}$ to order $n-1$ and
thus continues the calculation.
The result to fourth order is,
\begin{eqnarray}
&{\vec r}_{0}(t)
=
{1\over 2}{\dot{\vec v}}_{S}(0)t^{2}
+
{1\over {3!}}\left(
{\ddot{\vec v}}_{S}(0)
-
\tau^{-1}
{\dot{\vec v}}_{S}(0)
\right)
t^{3}\\
&+
{1\over {4!}}\left(
-\tau^{-1}\Omega_{0}[{\dot{\vec v}}_{S}(0)\times{\hat z}]
-\tau^{-1}{\ddot{\vec v}}_{S}(0)
+\tau^{-2}{\dot{\vec v}}_{S}(0)
\right)
t^{4}
+
\cdots
\end{eqnarray}
where $\Omega_{0}\equiv \Delta_{0}^{2}/\hbar E_{F}$.
This calculation is
fully consistent and equivalent to
taking the derivatives of equation \ref{MOTION}
and evaluating them at $t=0$. Since the equation is linear
and velocities can be calculated to all orders in $t$ this equation
is valid for all $t>0$.

Equation \ref{MOTION} at zero frequency was introduced by
deGennes and Matricon\cite{DEGENNES}. The dissipation
acts on ${\vec v}_{L}$ rather than ${\vec v}_{S}$
as in the NV equation.
This has the
drawback of not allowing for small
conductivities observed in experiments on
flux flow\cite{BORCHERDS} (not to mention that there
has never been a satisfactory explanation of the
Hall effect).
The present derivation is valid in the clean,
low temperature limit where the levels are clearly
separated. Our result agrees with the NV equation
of motion in that limit\cite{JING} but cannot be extended to the dirty
limit.

It is easily verified that the homogeneous solution of \ref{MOTION}
corresponds to circular motion, with a definite handedness, at
frequency $\Omega_{0}$ and decaying on a timescale $\tau$.
To obtain a prediction for the
surface impedance
consider the particular solution for
${\vec v}_{S}(t)=
{\vec v}_{S}(0)\exp{(i\omega t)}$.
It is
\begin{equation}
v_{Ly} - v_{Sy}
=
\left[
{{
\Omega_{0}\tau v_{Sx} - (1 + i\omega\tau)v_{Sy}
}\over{
(1 + i\omega\tau)^{2} + (\Omega_{0}\tau)^{2}
}}
\right],
\label{PARTICULAR}
\end{equation}
and another equation with x,y interchanged and
$\Omega_{0}\rightarrow -\Omega_{0}$.

There are two contributions to the surface impedance.
The first is transverse vortex motion
in phase with the supercurrent.
For clarity let $v_{Sy}=v_{Sx}\exp{i\theta}=v_{S}/\sqrt{2}$.
There is an induced voltage per vortex
${\hat z}\times {\vec v}_{L}(h/2e)$. The supercurrent density
$v_{S}ne$ gives dissipation
$N_{v}(hn/2)Re(v_{Lx}^{*}v_{Sy} - v_{Ly}^{*}v_{Sx})$,
where $N_{v}$ is the vortex density.
Using equation \ref{PARTICULAR}
\begin{eqnarray}
Re(v_{Lx}^{*}v_{Sy} - v_{Ly}^{*}v_{Sx}) &=
-v_{S}^{2}\Omega_{0}\tau
{
{1-(\omega\tau)^{2}+(\Omega_{0}\tau)^{2}}
\over
{[1-(\omega\tau)^{2}+(\Omega_{0}\tau)^{2}]^{2}+4(\omega\tau)^{2}}
}\\
&+ sin\theta v_{S}^{2}\omega\tau
{
{1+(\omega\tau)^{2}-(\Omega_{0}\tau)^{2}}
\over
{[1-(\omega\tau)^{2}+(\Omega_{0}\tau)^{2}]^{2}+4(\omega\tau)^{2}}
}.
\end{eqnarray}
The first term has an (anti)resonance at
$(\omega\tau)^{2}=(\Omega_{0}\tau)^{2}+1$
while the second, polarization dependent, term
has a resonance there.

The second source is the current due
to vortex motion which is in phase and parallel with the applied
electric field. A straightforward calculation gives the average
current density due to vortex motion
to be
$2v_{L}(k_{Fz}/\pi)(E_{F}/\Delta)^{2}N_{v}e$.
With the electric field $E=-(m/e){\dot {\vec v}}_{S}$ the
dissipation is
$-2m(k_{Fz}/\pi)(E_{F}/\Delta)^{2}N_{v}
Re(v_{Ly}^{*}i\omega v_{Sy} +
v_{Lx}^{*}i\omega v_{Sx})$.
Using equation \ref{PARTICULAR},
\begin{equation}
Re(v_{Ly}^{*}i\omega v_{Sy} +
v_{Lx}^{*}i\omega v_{Sx})
=
v_{S}^{2}
\omega^{2}\tau
{
{1+(\omega\tau)^{2}-(\Omega_{0}\tau)^{2}}
\over
{[1-(\omega\tau)^{2}+(\Omega_{0}\tau)^{2}]^{2}+4(\omega\tau)^{2}}
}.
\end{equation}
This has a peak at
$(\omega\tau)^{2}\sim(\Omega_{0}\tau)^{2}+1$
and is not polarization dependent.
By letting $\Omega_{0}\rightarrow 0$, this term
becomes simply the Drude expression
for dissipation.

The expected polarization dependent
absorption is distributed, due to states with different $k_{z}$,
over a range of frequencies. Precise details
depend on the fermi surface shape. The temperature dependence
should be weak.
The current due
to quasiparticles moving in and out of the vortex core
can be neglected at low temperature because there is
a discrete energy cost to make charge fluctuations in the core.
The considerations of this paper may be
valid even for pinned vortices.
If some parts of a line are pinned, other parts of
between pins
can move, provided they are excited at high
enough frequency.

The author was supported by the Killam foundation. He wishes
to acknowledge conversations with A.J. Berlinsky, S. Hagen, W. Hardy,
P. Schleger, A.-C. Shi and especially J. Rammer.

\end{document}